\documentclass[12pt]{iopart}

\newcommand{\be}{\begin{eqnarray}}
\newcommand{\ee}{\end{eqnarray}}
\newcommand{\ket}[1]{\ensuremath{\left| {#1} \right>}}
\newcommand{\bra}[1]{\ensuremath{\left< {#1} \right|}}

\newcommand{\dnpm}{\delta_n^{\pm}}
\newcommand{\upup}{\uparrow\uparrow}
\newcommand{\updown}{\uparrow\downarrow}
\newcommand{\downup}{\downarrow\uparrow}
\newcommand{\downdown}{\downarrow\downarrow}

\newcommand{\eqref}[1]{(\ref{#1})}

\usepackage{graphicx}
\usepackage{iopams}

\begin{document}

\title{Pulsed force sequences for fast phase-insensitive quantum gates in trapped ions}

\author{A M Steane$^1$, G Imreh$^1$, J P Home$^1$ and D Leibfried$^2$}

\address{$^1$ Clarendon Laboratory, Department of Physics, University of Oxford, Parks Road, Oxford OX1 3PU, UK}
\address{$^2$ NIST Boulder, Time and Frequency Division, Boulder Colorado 80305, USA}


\begin{abstract}
We show how to create quantum gates of arbitrary speed between trapped ions,
using a laser walking wave, with complete insensitivity to drift of the
optical phase, and requiring cooling only to the Lamb-Dicke regime. We present
pulse sequences that satisfy the requirements and are easy
to produce in the laboratory.
\end{abstract}



\maketitle

There exist a variety of proposals for quantum logic gates between
ions trapped in high vacuum
\cite{95Cirac,00Cirac,99Sorensen,00Sorensen1,00Milburn2,03GarciaRipoll,05GarciaRipoll,04Staanum,04Duan,03Beige,04Beige,00Jonathan,08Ospelkaus},
several of which have been experimentally implemented
\cite{98Turchette,00Sackett,03Leibfried,03SchmidtKaler,05Haffner,05Leibfried,06Home,05Brickman,11Ospelkaus,13Lanyon,12Hayes,13Mizrahi}.
For quantum computing, the primary goal is fast
precise quantum logic. Methods are sought
which are insensitive to those experimental parameters ${\cal P}_{\rm h}$ which
are, in practice, harder to control \cite{04Haljan,06Zhu,08Roos,12Bermudez,13Lemmer}. Such methods still require
accurate parameter settings, but only for parameters ${\cal P}_{\rm e}$ whose precise
control is available in the laboratory.
Examples of ${\cal P}_{\rm h}$ are
ion temperature $T$ and optical phases $\phi_{o}$; examples of ${\cal P}_{\rm e}$ are
r.f. frequencies and phases. An optical phase is the phase of a travelling- or
standing-wave of light which is sensitive to nm-scale changes in optical path lengths.
Single-qubit gate methods, such as stimulated Raman transitions, can use co-propagating
laser beams to avoid sensitivity to slow drift of $\phi_o$, and this is extremely
important to the realization of the high degree of precision which is needed
for quantum computing. Similar insensitivity is available in some
2-qubit gate methods, but at a cost in speed. We show how to construct
simple laser pulse sequences which allow arbitrarily fast gates
robust against $T$ and having complete
insensitivity to the value of the optical phase $\phi_o$ when the gate is
applied.

The issues of gate speed and sensitivity to $\phi_o$ have been addressed separately
in recent work. Using the general concept of forced displacements in
phase space \cite{00Sorensen1,00Milburn2,03Leibfried}, Garc\'ia-Ripoll {\em et al.}
\cite{03GarciaRipoll,05GarciaRipoll}
showed how to find the time-dependence $f(t)$ of a spin-dependent force which
would allow a 2-qubit gate of any speed in a given trap.
The issue of optical phases arises when the force is produced
by a laser standing wave: if $\phi_o$ were not controlled then the
ions would experience the wrong $f(t)$ and the gate would not work
in general.
For `slow' gates, on the other hand, i.e. with $\omega_c \tau \gg 2\pi$,
where $\omega_c$ is the centre of mass (COM) mode frequency, and
$\tau$ the gate time, one may use the simple laser pulse described
in \cite{03Leibfried} and below, resulting in an entangling gate
insensitive to $\phi_o$. The method of  M{\o}lmer and S{\o}rensen
involving spin flips \cite{99Sorensen,00Sorensen1}, as originally discovered, was
sensitive to $\phi_o$ even in the slow
regime, and this was an important limitation on its practical
realisability. However, it was shown how to choose
laser beam propagation directions so that
insensitivity to $\phi_o$ is possible for
a slow implementation of the gate on a pair of ions \cite{04Haljan}; see also \cite{12Hayes}.
Here we achieve both high speed and insensitivity to $\phi_o$.

The gate methods mentioned above work by
driving the system around
a closed loop in phase space, using an oscillating force whose magnitude
or direction depends on the qubit state. The closure
and the loop area can depend on the initial phase
of the force because for fast gates with
$\omega_c \tau \lesssim 2\pi$, the rotating wave approximation
is not valid. A given oscillation $\sin(\omega t + \phi_o)$
appears as an oscillation at two frequencies $\omega_0 \pm \omega$ in
an interaction picture. The phase $\phi_o$ is half the
phase difference between these two oscillations at $t=0$, and it
has significant physical consequences. Similar
considerations arise in fast manipulation of spins by Rabi flopping.

A summary of our analysis is as follows. We consider a pair
of ions subject to a laser walking wave, of phase $\phi_o$
at some arbitrary position and time, which produces a
spin-dependent oscillating force through the a.c. Stark shift.
We extract the dependence of the dynamics on $\phi_o$, and
obtain 4 complex number conditions which suffice to ensure
that all loops in the motional phase space are closed for
any value of $\phi_o$. We then extract a further 2 complex number
conditions which ensure that loop areas are independent of $\phi_o$,
and another condition which ensures that single-qubit rotations
due to a.c. Stark shifts also vanish for all $\phi_o$.
Next, we search for pulse sequences that satisfy all the
conditions. We show that some remarkably simple pulse sequences can
succeed at fast speeds $\omega_c\tau < 2\pi$: for
example, a symmetric 5-pulse sequence of fixed
frequency and phase origin (as could be produced by a single
oscillator gated between zero and three output levels). Such a
high speed requires high laser intensity, however, and this may
be impractical. We also
present a very simple symmetric 4-pulse sequence, with all pulses at
the same amplitude, which produces infidelity below $10^{-4}$ (averaged
over a uniform distribution of $\phi_o$) at gate
speed $\omega_c \tau/2\pi \simeq 1.76$.

Consider a walking wave of light interacting with two ions
in the same trap. The light field is produced by two laser beams
of difference wavevector $\bf k$ directed along $x$, difference
frequency and phase $\omega, \phi_o$. The internal qubit
states of the ions will be labelled $\ket{\uparrow}$ and $\ket{\downarrow}$
and referred to as `spin states'; in practice they are usually
a pair of levels in the ground state hyperfine structure. We assume
the light polarization is adjusted so that the a.c. Stark
shift from either laser beam acting alone
is the same for $\ket{\uparrow},\ket{\downarrow}$. This is usually
the case in experiments and serves to eliminate an error source
(random qubit rotations from laser intensity noise). The laser-ion
interaction Hamiltonian is then
\[
H = \sum_{m_1,m_2} \sum_j C_{m_j} {V}_j
\cos(k\hat{x}_j-\omega t -\phi_o - \varphi_{m_j} ) \ket{m}\bra{m}
\]
where $\ket{m}$ is shorthand for $\ket{m_1 m_2}$,
$j=0,1$ counts the ions, $m_j=\uparrow,\downarrow$ indicates
the internal state, the $C$ coefficients account for different
coupling to different spin states, $V_j(t)$ (real and positive) is
proportional to the product of the two laser beam electric field
amplitudes at ion $j$ at time $t$, and $\varphi_{m_j}$ depend
on the light polarization.

In general, for each two-ion spin state, both the centre of mass
(COM) and stretch modes of motion of the ions are excited, by
differing amounts. After
making the Lamb-Dicke approximation $|k q_j| \ll 1$, where $q_j =
x_j - x^0_j$ are the excursions from equilibrium, we have
\be
H & \simeq & \hbar \sum_{m} \ket{m}\bra{m} \left[ \rule{0 pt}{1.8 em}
\Omega^{\rm LS}_{m} \cos(\omega t + \phi_o + \phi^{\rm LS}_{m} ) \right.
\nonumber            \\
&& + \left.
 \sum_{l=c}^s \Omega_{m l}
\sin(\omega t  + \phi_o + \phi_{m l} )
k \hat{x}_l \right]          \label{HLD}
\ee
where $\{\Omega(t),\; \phi\}$ give the amplitudes and phases
of the various contributions, and
$x_c \equiv (q_1+q_2)/2$, $x_s \equiv (q_1-q_2)/2$
are the COM and stretch coordinates.
The equations relating $\Omega,\phi$
to $C,V,\varphi,k(x_2^0-x_1^0)$ are easily derived
but lengthy to write down.

The first term in (\ref{HLD}) is a time-dependent light shift (LS) (a.c. Stark shift)
causing spin precession about $z$. Although for slow gates it can be negligible, it
will be important here. The second term is a sum of time-dependent
forces acting on the normal modes. It is well known that the effect of such
a uniform force on quantum SHM is simply to displace the
motional state in its $x$--$p$ phase space \cite{65Carruthers,BkWalls}.
Let $\ket{\alpha,n}$ be a Fock state displaced by $\alpha$, then
the total evolution has the form
\be
\ket{m}\ket{0,{n}_c;\, 0,{n}_s} &\rightarrow& e^{i \Phi_m} \ket{m}
           \ket{\Delta\alpha_c^m, {n}_c;\, \Delta\alpha_s^m,{n}_s }.
\ee
The phases $\Phi_m$ have a contribution from the LS term,
and a contribution proportional to the sum of the (signed) areas
enclosed by the phase space orbits $\alpha_c^m(t), \alpha_s^m(t)$.
The desired 2-qubit phase gate is obtained when
$\Psi \equiv \Phi_{\upup}+\Phi_{\downdown}-\Phi_{\updown}-\Phi_{\downup} = \pi$.
$\Delta \alpha_{c,s}^m$ are the net displacements at the end of the gate
operation. In an ideal case these would be zero
so that the spin and motion are disentangled. We calculate the infidelity
$\epsilon = 1 - \left<\right. |\bra{\psi} U_{\rm id}^{\dagger} U \ket{\psi}|^2
\left. \right>$
where $U_{\rm id}$ is an ideal operation and the outer brackets represent averaging
over an initial thermal state, and over spin states \footnote{We average over
product spin states where each spin is uniformly distributed over the Block sphere.
The numerical factors in front of the terms in (\ref{eps})
depend on the averaging and on correlations amongst $\Phi_m$. However,
for solutions $\epsilon=0$ the values of these factors are irrelevant.}.
For small errors we obtain
\be
\epsilon &\simeq&
\sum_{l,m}
 \frac{ 1 + 2 \bar{n}_l }{4} |\Delta\alpha_l^m|^2
 + \frac{\Delta\Psi^2}{9} + \frac{ \Delta\theta_1^2 + \Delta\theta_2^2 }{5},
 \label{eps}
\ee
where
$\theta_{1,2} \equiv ((\Phi_{\upup}-\Phi_{\downdown})
\pm (\Phi_{\updown} - \Phi_{\downup}))/2$ are single-qubit rotation angles,
and $\Delta \Psi = \Psi - \pi$, $\Delta\theta_j = \theta_j - \bar{\theta_j}$

For an oscillator with mass $M$ and natural frequency $\omega_0$, the
coherent state parameter is defined as
$\alpha = \exp(i \omega_0 t) (x + i p/M\omega_0)/2 x_0$,
where $x_0 = (\hbar/2 M \omega_0)^{1/2}$.
The Argand diagram for $\alpha$ corresponds to an $x$--$p$ phase space in which the
motion can be conveniently described.
For a uniform driving
force $f(t)$, the evolution is given by \cite{65Carruthers,BkWalls}
 \be \Delta\alpha \equiv \alpha(t) - \alpha(0) =
\frac{i}{2 M \omega_0 x_0} \int_0^t e^{i \omega_0 t'} f(t') {\rm d}t'.
\label{at}
 \ee

Consider the force owing to any one of the terms in (\ref{HLD}):
\[
f(t) = 4 M \omega_0 x_0 \Omega(t) \sin(\omega t + \phi),
\]
where $\Omega = -\eta_l \Omega_{m l}/2$,
$\phi = \phi_o + \phi_{m l}$, with the Lamb-Dicke parameter
$\eta_l = k x_{0l}$. One finds
\be
\Delta \alpha = e^{i\phi} \Delta\alpha^+  +   e^{-i\phi} \Delta\alpha^-
\ee
where $\Delta\alpha^{\pm} = \pm \int_0^t \Omega(t') \exp(i \delta^{\pm} t') {\rm d}t'$
with $\delta^{\pm} \equiv \omega_0 \pm \omega$. Therefore $\Delta \alpha$
describes an ellipse in the Argand diagram as $\phi_o$ is varied. In order
to guarantee that $\Delta \alpha=0\, \forall \phi_o$, it is sufficient and necessary
that $|\Delta\alpha^{\pm}|=0$. If $\phi_o$
is uniformly distributed between $0$ and $2\pi$ then the mean value of
$|\Delta \alpha|^2$ is
$
\left< |\Delta \alpha|^2 \right> = |\Delta\alpha^+|^2 + |\Delta\alpha^-|^2.
$

When $\Delta\alpha=0$ the orbit is closed, and the phase $\Phi$ acquired
by the quantum state is twice the enclosed area, $\Phi = {\rm Im}[I]$ where
\be
I &=& \int_{\rm path} \alpha^* {\rm d}\alpha
= I^0 + e^{2 i \phi} I^+ + e^{-2 i \phi} I^-   \label{I}
\ee
with
\be
I^0 &=& \int_0^t \Omega(t') \left(
\Delta\alpha^{+*} e^{i \delta^+ t'} - \Delta\alpha^{-*} e^{-i\delta^- t'} \right) {\rm d}t' \\
I^{\pm} &=& \pm \int_0^t \Omega(t') \Delta\alpha^{\mp *}(t') e^{i \delta^{\pm} t'} {\rm d}t'.
\ee
In order that $\Phi$ is independent of $\phi_o$ it is sufficient and necessary
that $I^+ = I^{-*}$.
When $\phi_o$ is uniformly distributed between $0$ and $2\pi$,
the variance of $\Phi$ is
$
\Delta\Phi^2 = |I^+ - I^{-*}|^2 / 2.
$

Now consider the $\Omega^{\rm LS}_m$ term in \eqref{HLD}.
The contribution to $\Phi_m$ is
$\theta_m^{\rm LS} \equiv \exp(i \phi_o) \theta_m^{+} + {\rm c.c.}$ where
$\theta_m^{+} =
- \int (\Omega_m^{\rm LS}/2) \exp i(\omega t + \phi_m^{\rm LS}) {\rm d}t$.
This does not contribute to $\Psi$ but produces single-qubit rotations.
When $\phi_o$ is uniformly distributed, $\theta_m^{\rm LS}$ has mean zero and variance
$2|\theta_m^{+}|^2$.

Now consider a sequence of laser pulses, where in general
the amplitude, frequency and relative phases of the pulses
may differ, although later we will restrict to all $\omega_n$
and $\phi_n$ the same. The force on a given
mode for a given spin state has the form
\be f(t) = \sum_{n=1}^N T\left(
({t-t_n})/{\tau_n} \right) f_n \sin(\omega_n t + \phi_n),
\label{ft} \ee
where the `top hat' function $T(x)$ is 1 for $0 < x \le
1$ and zero otherwise. Thus the $n$'th pulse begins at $t_n$ and
has duration $\tau_n$. Then during any given pulse
the change in $\alpha$ is given by
$
\Delta \alpha_n (t,\phi_o) = A_n^+(t) e^{i\phi_o} + A_n^-(t) e^{-i\phi_o}
$
where
\be
A_n^{\pm}(t) = {\pm i\Omega_n}
              e^{i(\dnpm t_n \pm \Delta \phi_n)}
              \left(1-e^{i\dnpm (t-t_n)} \right) / {\dnpm},  \label{Anpm}
\ee
and $\Delta \phi_n \equiv \phi_n - \phi_o$.
$\Delta\alpha_n(t,\phi_o)$ describes a cycloid.

Let $A_n^{\pm} \equiv A_n^{\pm}(\tau_n)$. The orbit area calculation (\ref{I}) gives
\be
I^0 &=& \sum_n \alpha_n^{+*} A_n^+ + \alpha_n^{-*} A_n^- + B_n^0,  \\
I^{\pm} &=& \sum_n \alpha_n^{\mp *} A_n^{\pm}  + B_n^{\pm}
\ee
where
\be
\alpha_n^{\pm} &=& \sum_{j=1}^{n-1} A_j^{\pm},  \\
B_n^0 &=& \Omega_n^2\left[ \frac{2i\omega_0\tau_n}{\delta_n^+ \delta_n^-}
+ \frac{C_n(\delta_n^+)}{\delta_n^+}
+ \frac{C_n(\delta_n^-)}{\delta_n^-} \right],  \\
B_n^{\pm} &=& {\Omega_n^2} e^{\pm 2 i \tilde{\phi}_n }
\left( C_n(\pm 2\omega_n) - C_n(\dnpm) \right)/{ \delta_n^{\mp} }.
\ee
For brevity we introduced the circle function $C_n(\omega) \equiv (1-\exp(i\omega \tau_n))/\omega$,
and phase $\tilde{\phi}_n = \omega_n t_n + \Delta\phi_n$.
The LS term gives
\be
\theta_m^+ = \sum_n (-i \Omega_{m,n}^{\rm LS}/2) \exp(i \tilde{\phi}_n) C_n(\omega_n).
\label{LSterm}
\ee

We are interested in the case where $\Delta\phi_n$ are well-defined but
$\phi_o$ is not.
Our problem is to find a sequence of pulses such that $\epsilon \ll 1$
when $\phi_o$ is uncontrolled,
with the total time $\tau \equiv t_N+\tau_N-t_1$ small. Smaller pulse
magnitudes and areas are preferred, to minimize the laser intensity
and decoherence from photon scattering. Also the number
of parameters describing the sequence should be small, to reduce the control
problem.

To solve for general values of
the coupling coefficients, it is sufficient to find a sequence producing
\be
\Delta\alpha_c^+ &=& \Delta\alpha_c^- = \Delta\alpha_s^+ = \Delta\alpha_s^- = 0, \label{alpha0} \\
\theta^+ &=& 0,  \label{theta0} \\
I_c^+ &=& I_c^{-*}, \; I_s^+ = I_s^{-*},
\ee
for non-zero $\Omega_{ml}$, because orbits of different spin states only differ by
an amplitude factor and phase origin. Therefore there
are 7 complex numbers that must be zero.
By contrast, if we only needed a solution at one value of $\phi_o$, there would
be only two complex number conditions: $\Delta\alpha_c = 0,$ $\Delta\alpha_s = 0$.
One can drop the condition on the LS term $\theta^+$ while doubling the total
gate time, by applying a given pulse sequence twice, with a spin-flip in
between (spin-echo sequence). However $\theta^+$ must not be ignored altogether
because for fast gates the LS phases are greater than the orbit area
phases by approximately $1/\eta$.

\begin{figure}[ht!]
\centerline{\resizebox{!}{0.4\textwidth}{\includegraphics{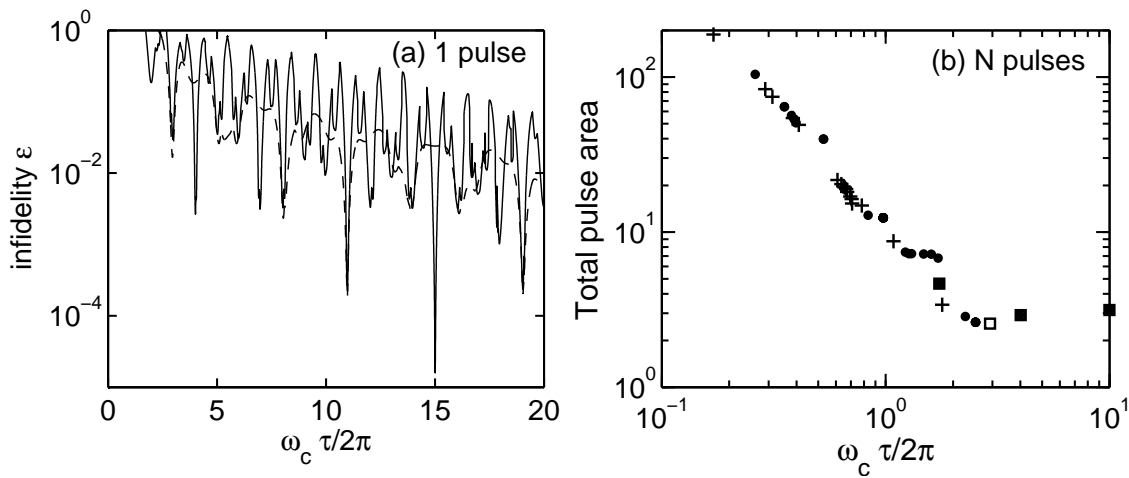}}}
\caption{(a) Infidelity verses $\omega_c \tau/2\pi$ for a gate using a single laser pulse.
The case $\eta_c = 0.1$, $\{ \bar{n}_c, \bar{n}_s \} = \{1,1\}$ is shown,
for $C_{\uparrow}=-C_{\downarrow}$, $V_1 = V_2$, $k(x_2^0-x_1^0)=2\pi p$
(equal and opposite light shifts, balanced intensities, ion separation an integer
number $p$ of standing wave periods). At each $\tau$, $\omega$ is optimized.
Full curve: single pulse, dashed curve: spin-echo sequence, with
two pulses of duration $\tau/2$ and $\theta^+ \ne 0.$
(b) Near-exact solutions: total pulse area vs. total gate time, for
example pulse sequences. The examples shown are time-symmetric with
fixed $\omega_n=\omega$, $\Delta\phi_n=0$.
$+,\centerdot,\square,\blacksquare$: $N=6,5,4,4$ pulses;
$\blacksquare$ has all pulses of same amplitude.
}
\end{figure}

For each pulse, a change in
$\phi_o$ rotates and displaces the orbit along itself:
$\Delta\alpha_n(t,\phi_o) = \exp(-i\theta_d) (\Delta\alpha_n(t+t_d,0)-\Delta\alpha_n(t_n+t_d,0))$
where $\theta_d=\phi_o \omega_0/\omega_n$, $t_d=\phi_o/\omega_n$.
Therefore if it were possible to close both orbits $\alpha_l$ with a single
pulse, then the closure would be guaranteed for all $\phi_o$, and also the areas
would be independent of $\phi_o$. However, because $\omega_s/\omega_c = \sqrt{3}$
is irrational, this is not possible\footnote{In an anharmonic trap one can have
$\omega_s/\omega_c$ rational, but then the mode frequencies and the ion separation
become sensitive to stray d.c. electric fields.}. We show in figure 1a the
best that can be done with a single pulse. There are two free parameters
$\omega$, $\tau$. The minimum infidelity $\min_{\omega} \epsilon$
was calculated for durations in the range $(1-20)/2\pi\omega_c$, at $\eta_c=0.1$,
$\bar{n}_c = \bar{n}_s = 1$. It is seen that high fidelity can be obtained when
$\omega_c \tau/2\pi$ is near to the denominator of a rational approximation to
$\sqrt{3}$ (i.e. the values 4,11,15), but $\epsilon \le 10^{-4}$ is not
available for $\omega_c \tau/2\pi < 15$.

We performed a numerical search for fast pulse sequences which solve the problem.
A sequence was deemed a `solution'\footnote{When $\epsilon \ll 10^{-4}$ in \eqref{eps}, the
fidelity will be limited in practice by other considerations, such as breakdown of
the Lamb-Dicke approximation or laser intensity noise.} if $\epsilon < 10^{-8}$
at $\eta_c = 0.1$, $\bar{n}_c = \bar{n}_s = 1$, or if $\epsilon < 3 \times 10^{-5}$ 
in the case of a pulse sequence with fewer than seven parameters.
A sequence of $N$ pulses has $5N-3$ parameters,
since the start time and phase origin are arbitrary, and the absolute
amplitude is fixed by the requirement $\Psi = \pi$. This suggests
that solutions might be possible with few pulses. However
to simplify experimental requirements we assumed fixed $\omega_n = \omega$
and restricted the values of $\Delta\phi_n$. For example, useful solutions were found with
$\Delta\phi_n$ restricted to multiples of $\pi/2$, and also with $\Delta\phi_n = 0$.
For $\Delta\phi_n=0$ and fixed $\omega_n$
the number of parameters is $3N-1$.
We found that by further restricting to time-symmetric
sequences ($\lceil 3N/2 \rceil$ parameters),
the rapidity with which solutions were found increased. This is because
the parameter space is smaller, but it shows the symmetric space
contains a good density of solutions. Another possibility is to remove
the gaps between the pulses, so that the sequence describes a single
shaped pulse, with $2N$ parameters if all sections of the pulse 
have the same frequency and phase origin. We find there are
solutions at $\omega_c \tau /2\pi \simeq 2.93$ 
for a shaped pulse with three sections, and symmetric shaped pulses with
5 sections can give faster solutions (see for example figure \ref{fig3}).

\begin{figure}[ht!]
\centerline{\resizebox{!}{0.5\textwidth}{\includegraphics{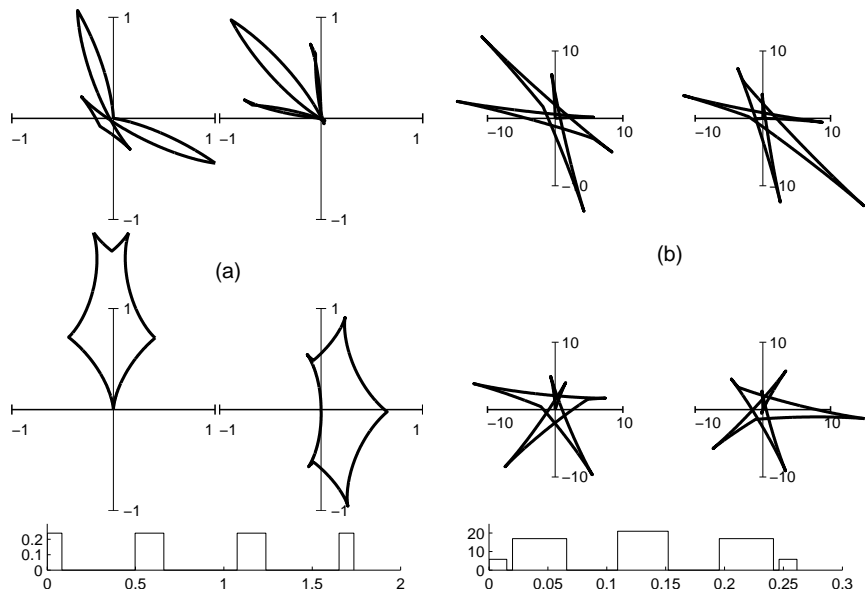}}}
\caption{Behaviour for two example symmetric pulse sequences:
(a) moderately fast and simple, (b) very fast.
The orbit $\alpha$ is shown for COM (top) and stretch (bottom) modes,
for two values of $\phi_o=0$ (left) and $\pi/2$ (right), with
pulse amplitude vs. $t$ at the bottom, in units of the COM period, $2\pi/\omega_c$.
Case (a) has $\omega\simeq 4.0376 \omega_c$,
pulse durations and gaps
$\tau_1, t_2-\tau_1, \tau_2, \ldots \simeq$
$\{0.524696 ,\, 2.60288 ,\, 1.02264 ,\, 2.60407 \}/\omega_c$.
Case (b) has
$\omega\simeq 20.4761  \omega_c$,
$\tau_1, t_2-\tau_1, \tau_2, \ldots, \tau_3 =
\{ 0.0953071$, $0.0305622$, $0.288972$, $0.272151$, $0.269998 \}/\omega_c,$
relative pulse amplitudes $\Omega_{1,2,3} = \{ 1,\,   2.91057,\, 3.59685\}$.
}
\label{fig2}
\end{figure}

\begin{figure}[ht!]
\centerline{\resizebox{!}{0.5\textwidth}{\includegraphics{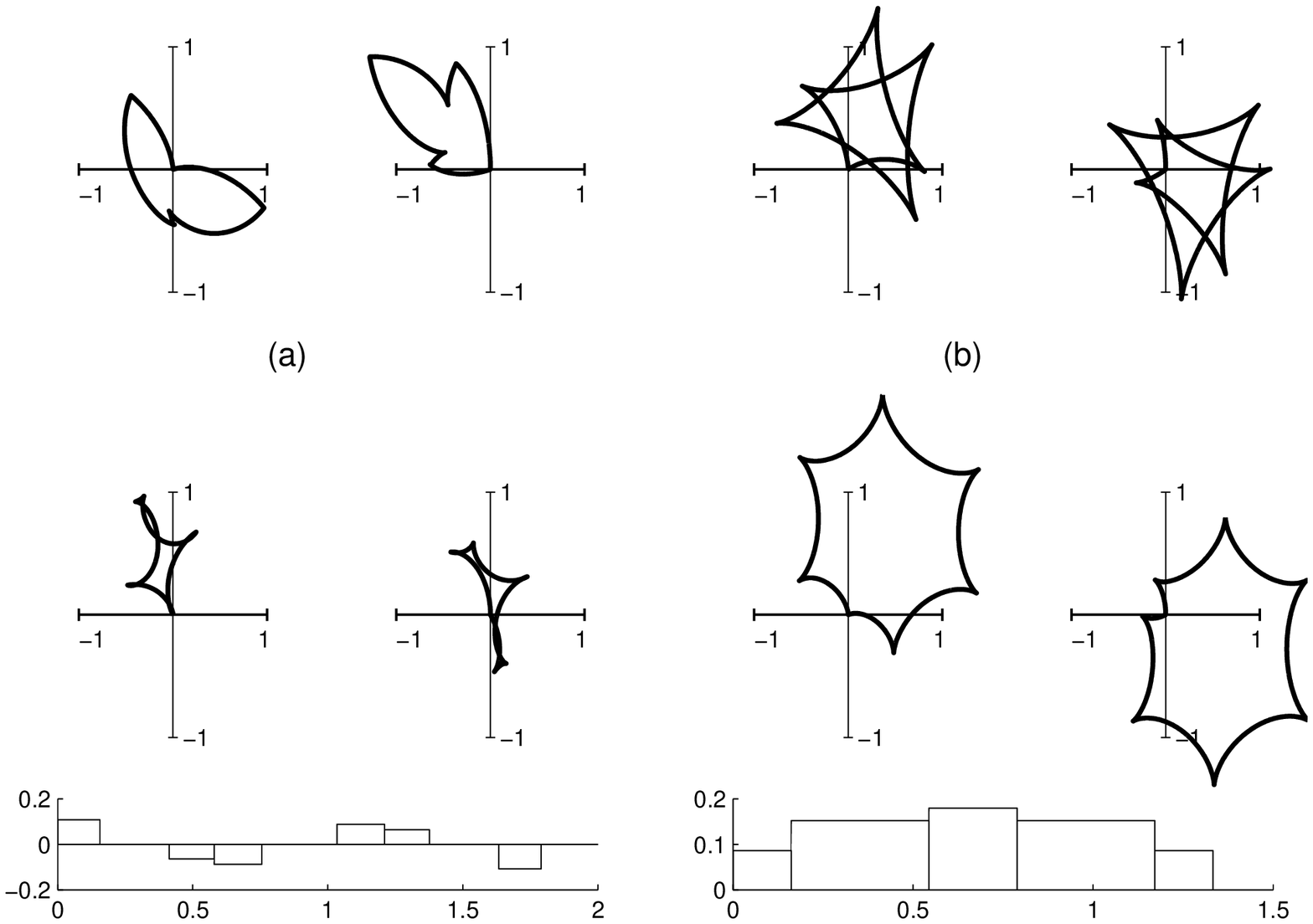}}}
\caption{Behaviour for two further examples, both moderately fast and simple.
(a) has a small total pulse area, (b) illustrates a single shaped pulse.
Case (a) has $\omega\simeq 1.36603 \omega_c$,
pulse durations and gaps
$\tau_1, t_2-\tau_1, \tau_2, \ldots \simeq$
$\{0.984464,\,  1.6124,\,  1.04219,\,  0.0003\, 1.10990,\, 1.7475  \}/\omega_c$;
pulse amplitudes $\Omega_{1,2,3}=$ $\{0.6786,\,   -0.4002,\,   -0.5528\}$.
Case (b) has $\omega\simeq 2.60258  \omega_c$, segment durations $\tau_{1,2,3}=$
$\{  1.0168,\,    2.3997,\,    1.5416 \}/\omega_c,$
pulse amplitudes $\Omega_{1,2,3} = \{   0.5415 ,\,   0.9561  ,\,  1.1280   \}$.
}
\label{fig3}
\end{figure}

To find a solution, we searched among values of $\omega,t_n,\tau_n$
from a random starting point, for each case solving a set of linear
equations for the $N-1$ amplitudes $\Omega_{n>1}$, and using the
Nelder-Mead simplex method and simulated annealing to find a minimum
of $\epsilon$. The linear equations were a subset of (\ref{theta0}),
(\ref{alpha0}). We found that $N=4$ can give some quite fast pulse
sequences ($\tau \simeq 4\pi / \omega_c$) at $\epsilon =
O(10^{-5})$, see figure \ref{fig2}, and $N=5$ was sufficient and necessary to
get fast solutions ($\tau < 2\pi/\omega_c$, $\epsilon < 10^{-8}$).
In figure 1b we show the total pulse area $\int \Omega(t) {\rm d}t$ versus
length of the pulse sequence, for sequences of minimal area at given
$\tau$. The area is important because the unwanted photon scattering
is proportional to it. With a single (slow) pulse the area is
approximately $\pi$. For $\omega_c \tau /2\pi < 2$ we find the same
$\tau^{-3/2}$ scaling law as was described in \cite{05GarciaRipoll}.

The terms in (\ref{LSterm}) contribute single-qubit phases that are each
of order $\pi/(N\eta)$. They cancel when the frequency and timing are accurate, but
in the presence of timing errors these are the main source of infidelity. However their
influence can be reduced by employing a spin-echo, and as long as any timing
error is constant, the value of $\theta_m^+$ can be adjusted to high accuracy
in practice by tweaking the pulse frequency $\omega$. Once this is done,
the dependence of infidelity $\epsilon$ on the fractional inaccuracy $\sigma = \Delta p/p$
of other parameters $p$ (such as pulse duration or height) is
of order $\epsilon \sim 10 \sigma^2$.

For a fast gate the orbits have to be large to enable the required
phase difference to be acquired rapidly (cf. figure 2b). Eventually the motion goes
outside the Lamb-Dicke regime and then eqs (\ref{HLD}) to
(\ref{Anpm}) must be replaced by a more general analysis or
numerical integration.

A fast gate with low photon scattering leads to high laser power
requirements, and this may in practice limit what speed one would aim to achieve.
Pulse shapes will not be exactly square, but small
amounts of rounding can be accommodated by small adjustments to the
parameters. Overall, we have achieved a simple, practical
solution which offers to increase gate speed by an order of
magnitude, while maintaining insensitivity to optical phase drift,
with little cost in photon scattering rates.

The gate method we have analyzed has the disadvantage that the qubits must be stored, for the duration
of the gate, in levels whose energy difference has a first-order Zeeman effect. Supposing the qubits are
ordinarily in `clock' states, this can easily be achieved
by fast microwave pulses before and after the gate, but it is natural to enquire whether this
aspect can be avoided altogether, for example by exploiting the M{\o}lmer-S{\o}rensen
gate \cite{99Sorensen,00Sorensen1}. The latter does not require a first order Zeeman effect.
However, for running-wave-driven M{\o}lmer-S{\o}rensen gates, carrier transitions are only detuned on the order of the motional frequencies and cannot be neglected for the timescales anticipated in our method. The carrier terms do not commute with the sideband-tems under the M{\o}lmer-S{\o}rensen interaction, therefore it is not straightforward to integrate the equations of motion \cite{00Sorensen1}. The situation is different for M{\o}lmer-S{\o}rensen gates induced by standing waves or microwave near-fields where the fields can be designed so that the carrier term, which is proportional to the field strength, vanishes to first order and also commutes with the sideband terms \cite{08Ospelkaus,11Ospelkaus,13Allcock}. 
Still, to produce standing wave fields with precise indexing to the position of single ions and high enough microwave gradients to drive sidebands with Rabi-frequencies sufficient for our method is very challenging with existing technology. It would
be interesting to discover whether the simpler strategy of allowing a Zeeman effect during the gate
would in fact be acceptable in the lab, because the gate is fast enough to make the accumulated phase owing
to a magnetic field fluctuation sufficiently small.

We thank D. M. Lucas and D. J. Szwer for useful discussions. This work was supported by
the National Security Agency (NSA) and Disruptive Technology
Office (DTO) (W911NF-05-1-0297), EPSRC (QIP IRC) and the Royal Society.

\bibliographystyle{unsrt}
\bibliography{myrefs}

\begin{thebibliography}{10}

\bibitem{95Cirac}
J.~I. Cirac and P.~Zoller.
\newblock Quantum computations with cold trapped ions.
\newblock {\em Phys. Rev. Lett.}, 74(20):4091--4094, 1995.

\bibitem{00Cirac}
J.~I. Cirac and P.~Zoller.
\newblock A scalable quantum computer with ions in an array of microtraps.
\newblock {\em Nature}, 404:579--581, 2000.

\bibitem{99Sorensen}
A.~S{\o}rensen and K.~M{\o}lmer.
\newblock Quantum computation with ions in thermal motion.
\newblock {\em Phys. Rev. Lett.}, 82:1971--1974, 1999.

\bibitem{00Sorensen1}
A.~S{\o}rensen and K.~M{\o}lmer.
\newblock Entanglement and quantum computation with ions in thermal motion.
\newblock {\em Phys. Rev. A}, 62:022311, 2000.
\newblock quant-ph/0002024.

\bibitem{00Milburn2}
G.~J. Milburn, S.~Schneider, and D.~F. James.
\newblock Ion trap quantum computing with warm ions.
\newblock {\em Fortschr. Physik}, 48:801--810, 2000.

\bibitem{03GarciaRipoll}
J.~J. Garcia-Ripoll, P.~Zoller, and J.~I. Cirac.
\newblock Speed optimized two-qubit gates with laser coherent control
  techniques for ion trap quantum computing.
\newblock {\em Phys. Rev. Lett.}, 91:157901, 2003.

\bibitem{05GarciaRipoll}
J.~J. Garcia-Ripoll, P.~Zoller, and J.~I. Cirac.
\newblock Coherent control of trapped ions using off-resonant lasers.
\newblock {\em Phys. Rev. A}, 71:062309, 2005.

\bibitem{04Staanum}
Peter Staanum, Michael Drewsen, and Klaus Moelmer.
\newblock Geometric quantum gate for trapped ions based on optical dipole
  forces induced by gaussian laser beams.
\newblock {\em Phys. Rev. A}, 70:052327, 2004.
\newblock quant-ph/0406186.

\bibitem{04Duan}
L.-M. Duan, B.~B. Blinov, D.~L. Moehring, and C.~Monroe.
\newblock Scalable trapped ion quantum computation with a probabilistic
  ion-photon mapping.
\newblock 2004.
\newblock quant-ph/0401020.

\bibitem{03Beige}
A.~Beige.
\newblock Dissipation-assisted quantum gates with cold trapped ions.
\newblock {\em Phys. Rev. A}, 67:020301, 2003.

\bibitem{04Beige}
Almut Beige.
\newblock Ion-trap quantum computing in the presence of cooling.
\newblock {\em Phys. Rev. A}, 69:012303, 2004.

\bibitem{00Jonathan}
D.~Jonathan, M.~B. Plenio, and P.~L. Knight.
\newblock Fast quantum gates for cold trapped ions.
\newblock {\em Phys. Rev. A}, 62:042307, 2000.
\newblock quant-ph/0002092.

\bibitem{08Ospelkaus}
C.~Ospelkaus, C.~E. Langer, J.~M. Amini, K.~R. Brown, D.~Leibfried, and D.~J.
  Wineland.
\newblock Trapped-ion quantum logic gates based on oscillating magnetic fields.
\newblock {\em Phys. Rev. Lett.}, 101:090502, Aug 2008.

\bibitem{98Turchette}
Q.~A. Turchette, C.~S. Wood, B.~E. King, C.~J. Myatt, D.~Leibfried, W.~M.
  Itano, C.~Monroe, and D.~J. Wineland.
\newblock Deterministic entanglement of two trapped ions.
\newblock {\em Phys. Rev. Lett.}, 81(17):3631--3634, 1998.

\bibitem{00Sackett}
C.~A. Sackett, D.~Kielpinski, B.~E. King, C.~Langer, V.~Meyer, C.~J. Myatt,
  M.~Rowe, Q.~A.~Turchette amd W.~M.~Itano, D.~J. Wineland, and C.~Monroe.
\newblock Experimental entanglement of four particles.
\newblock {\em Nature}, 404:256, 2000.

\bibitem{03Leibfried}
D.~Leibfried, B.~DeMarco, V.~Meyer, D.~Lucas, M.~Barrett, J.~Britton, W.~M.
  Itano, B.~Jelenkovic, C.~Langer, T.~Rosenband, and D.~J. Wineland.
\newblock Experimental demonstration of a robust, high-fidelity geometric two
  ion-qubit phase gate.
\newblock {\em Nature}, 422:412--415, 2003.

\bibitem{03SchmidtKaler}
Ferdinand Schmidt-Kaler, Hartmut H�ffner, Mark Riebe, Stephan Gulde, Gavin
  P.~T. Lancaster, Thomas Deuschle, Christoph Becher, Christian~F. Roos,
  J�rgen Eschner, and Rainer Blatt.
\newblock Realization of the cirac-zoller controlled-not quantum gate.
\newblock {\em Nature}, 422:408--411, 2003.

\bibitem{05Haffner}
H.~H\"{a}ffner, W.~H\"{a}nsel, C.~F. Roos, J.~Benhelm, D.~Chek{-}al{-}kar,
  M.~Chwalla, T.~K\"{o}rber, U.~D. Rapol, M.~Riebe, P.~O. Schmidt, C.~Becher,
  O.~G\"{u}hne, W.~D\"{u}r, and R.~Blatt.
\newblock Scalable multiparticle entanglement of trapped ions.
\newblock {\em Nature}, 438:643, 2005.

\bibitem{05Leibfried}
D.~Leibfried, E.~Knill, S.~Seidelin, J.~Britton, R.~B. Blakestad,
  J.~Chiaverini, D.~B. Hume, W.~M. Itano, J.~D. Jost, C.~Langer, R.~Ozeri,
  R.~Reichle, and D.~J. Wineland.
\newblock Creation of a six-atom `{S}chr\"odinger cat' state.
\newblock {\em Nature}, 438:639--642, 2005.

\bibitem{06Home}
J.~P. Home, M.~J. McDonnell, D.~M. Lucas, G.~Imreh, B.~C. Keitch, D.~J. Szwer,
  N.~R. Thomas, S.~C. Webster, D.~N. Stacey, and A.~M. Steane.
\newblock Deterministic entanglement and tomography of ion spin qubits.
\newblock {\em New J. Phys.}, 2006.
\newblock quant-ph/0603273.

\bibitem{05Brickman}
K.-A. Brickman, P.~C. Haljan, P.~J. Lee, M.~Acton, L.~Deslauriers, and
  C.~Monroe.
\newblock Implementation of {G}rover's quantum search algorithm in a scalable
  system.
\newblock {\em Phys. Rev. A}, 72:050306(R), 2005.

\bibitem{11Ospelkaus}
C.~Ospelkaus, U.~Warring, Y.~Colombe, K.~R. Brown, J.~M. Amini, D.~Leibfried,
  and D.~J. Wineland.
\newblock Microwave quantum logic gates for trapped ions.
\newblock {\em Nature}, 476:181–184, 2011.
\newblock doi:10.1038/nature10290.

\bibitem{13Lanyon}
B.~P. Lanyon, P.~Jurcevic, M.~Zwerger, C.~Hempel, E.~A. Martinez, W.~D\"ur,
  H.~J. Briegel, R.~Blatt, and C.~F. Roos.
\newblock Measurement-based quantum computation with trapped ions.
\newblock {\em Phys. Rev. Lett.}, 111:210501, Nov 2013.

\bibitem{12Hayes}
D.~Hayes, S.~M. Clark, S.~Debnath, D.~Hucul, I.~V. Inlek, K.~W. Lee,
  Q.~Quraishi, and C.~Monroe.
\newblock Coherent error suppression in multiqubit entangling gates.
\newblock {\em Phys. Rev. Lett.}, 109:020503, Jul 2012.

\bibitem{13Mizrahi}
J.~Mizrahi, C.~Senko, B.~Neyenhuis, K.~G. Johnson, W.~C. Campbell, C.~W.~S.
  Conover, and C.~Monroe.
\newblock Ultrafast spin-motion entanglement and interferometry with a single
  atom.
\newblock {\em Phys. Rev. Lett.}, 110:203001, May 2013.

\bibitem{04Haljan}
P.~C. Haljan, K.-A. Brickman, L.~Deslauriers, P.~J. Lee, and C.~Monroe.
\newblock Spin-dependent forces on trapped ions for phase-stable quantum gates
  and motional schr\"{o}dinger cat states.
\newblock 2004.
\newblock quant-ph/0411068.

\bibitem{06Zhu}
S.-L. Zhu, C.~Monroe, and L.-M. Duan.
\newblock Arbitrary-speed quantum gates within large ion crystals through
  minimum control of laser beams.
\newblock {\em Europhys. Lett.}, 73:485--491, 2006.

\bibitem{08Roos}
C.~Roos.
\newblock Ion trap quantum gates with amplitude-modulated laser beams.
\newblock {\em New J. Phys.}, 10:013002, 2008.

\bibitem{12Bermudez}
A.~Bermudez, P.~O. Schmidt, M.~B. Plenio, and A.~Retzker.
\newblock Robust trapped-ion quantum logic gates by continuous dynamical
  decoupling.
\newblock {\em Phys. Rev. A}, 85:040302, Apr 2012.

\bibitem{13Lemmer}
A.~Lemmer, A.~Bermudez, and M.~B. Plenio.
\newblock Driven geometric phase gates with trapped ions.
\newblock {\em New J. Phys.}, 15:083001, 2013.

\bibitem{65Carruthers}
P.~Carruthers and M.~M. Nieto.
\newblock Coherent states and the forced quantum oscillator.
\newblock {\em Am. J. Phys.}, 7:537--544, 1965.

\bibitem{BkWalls}
D.~F. Walls and G.~J. Milburn.
\newblock {\em Quantum Optics}.
\newblock Springer, Berlin, 1994.

\bibitem{13Allcock}
D.~T.~C. Allcock, T.~P. Harty, C.~J. Ballance, B.~C. Keitch, N.~M. Linke, D.~N.
  Stacey, and D.~M. Lucas.
\newblock A microfabricated ion trap with integrated microwave circuitry.
\newblock {\em Applied Physics Letters}, 102:044103, 2013.

\end{thebibliography}

\end{document}